\documentstyle[twoside,fleqn,espcrc2]{article}
\title{N* Mass Spectrum from an Anisotropic Action}
\author{Frank X. Lee
\address{Center for Nuclear Studies, Department of Physics,
     The George Washington University,  Washington, DC 20052, USA}
\address{Jefferson Lab, 12000 Jefferson Avenue, Newport News, VA 23606, USA}
\thanks{Lattice Hadron Physics Collaboration}}
\begin{document}
\begin{abstract}
Results are reported for N* masses in the $1/2+$ and $1/2-$ sectors
on an anisotropic lattice.
The gauge action is the usual plaquette plus rectangle type,
the quark action is of the D234 type, 
both having tadpole-improved tree-level coefficients.
Clear splittings from the nucleon ground state are observed with smeared 
operators and 500 configurations.
The first excited states in each channel are further isolated,
using a correlation matrix and the variational method.
The basic pattern of these splittings is consistent with 
experimental observations.
\end{abstract}
\maketitle

\section{Introduction}
There is increasing experimental information on the N* spectrum 
from JLab and other accelerators, and the associated desire 
to understand it from first principles.
Given that state-of-the-art lattice QCD simulations have produced a ground-state
spectrum that is very close to the observed values~\cite{cppacs00},
it is important to extend beyond the ground states.
The rich structure of the N* spectrum~\cite{pdg00}
provides a fertile ground for exploring how the internal 
degrees of freedom in the nucleon are excited 
and how QCD works in a wider context.
One outstanding example is the splitting pattern
between the observed $1/2+$ spectrum and its parity partner: the $1/2-$ spectrum.
The splittings are a direct manifestation of spontaneous chiral symmetry breaking 
of QCD, because without it QCD predicts parity doubling in the baryon spectrum.

Several lattice studies of the N* spectrum have appeared recently.
In~\cite{lee99} the $1/2\pm$ spectrum was first explored with 
a ${\cal O}(a^2)$ improved next-nearest-neighbor action.
In~\cite{sas00} the splitting $N_{1/2-}-N_{1/2+}$ was examined 
with the domain-wall fermion action.
In~\cite{dgr00} results are reported for the $N_{1/2\pm}$ and 
the $\Delta_{3/2\pm}$ based on the non-perturbative ${\cal O}(a)$ 
improved clover action.
All the calculations show a clear splitting of the $N_{1/2-}$ 
from the ground-state nucleon.
In this work, we report further calculations in the $1/2\pm$ sectors
using a highly-improved action on a anisotropic lattice.  

\section{Method}
Two independent, local interpolating fields coupling to the nucleon 
are considered:
\begin{equation}
\chi_1(x) = \epsilon^{abc}
                 \left ( u^{Ta}(x) C \gamma_5 d^b(x) \right ) u^c(x) \, ,
\end{equation}
\begin{equation}
\chi_2(x) = \epsilon^{abc}
                 \left ( u^{Ta}(x) C d^b(x) \right ) \gamma_5 u^c(x) \, .
\end{equation}
The interpolating fields for other members of the octet can be found
by appropriate substitutions of quark fields~\cite{lein91}.
Despite having explicit positive-parity by construction, these interpolating
fields couple to both positive and negative parity states.
A parity projection is needed to separate the two.
In the large Euclidean time limit, 
the correlator with Dirichlet boundary condition in the time direction
and zero spatial momentum becomes
\begin{equation}
G(t) = \sum_{\bf x} <0|\chi(x)\,\bar{\chi(0)}|0>
\end{equation}
\begin{equation}
= \lambda_+^2 { \gamma_4 +1 \over 2} e^{- M_+ \, t}
+ \lambda_-^2 {-\gamma_4 +1 \over 2} e^{- M_- \, t}
\end{equation}
The relative sign in front of $\gamma_4$ provides the solution: taking the
trace of G(t) with $(1\pm\gamma_4)/4$ respectively. 
A method on how to do the projection at finite momentum was also proposed 
in~\cite{lee99} which involves only diagonal elements.
Three types of correlators are considered in this work: two diagonal 
ones $N_1=<\chi_1\bar{\chi_1}>$ and $N_2=<\chi_2\bar{\chi_2}>$,
and the crossed $N_3=<\chi_2\bar{\chi_1}> + <\chi_1\bar{\chi_2}>$.

The anisotropic gauge action of~\cite{mor97}, and 
the anisotropic D234 quark action of~\cite{alf98} are used.
Both have tadpole-improved tree-level coefficients.
Similar anisotropic actions have been used to study 
systems with heavy quarks~\cite{lewis00,wolo00,rbc00}.
One advantage of anisotropic lattices is that with modest lattice sizes 
one can access large spatial volumes while having a fine temporal resolution,
important when the states are extended and heavy.
A $10^3\times 30$ lattice with $a_s\approx 0.24$ fm 
and anisotropy $\xi=a_s/a_t=3$ is used. 
The tadpole factor from the plaquette mean link is $u_s=0.81$ while
$u_t$ is set to 1. In all, 500 configurations are analyzed. 
On each configuration 
9 quark propagators are computed using a multi-mass solver, 
with quark masses ranging from approximately
780 to 90 MeV. The corresponding mass ratio $\pi/\rho$ is from 0.93 to 0.60.
The source is located at the 2nd time slice.
A gaussian-shaped, gauge-invariant
smearing function~\cite{Gus90} was applied at both the source and the sink 
to increase the overlap with the states in question.
Statistical errors are derived from a bootstrap procedure.

\section{Results and Discussion}

\begin{figure}
\includegraphics{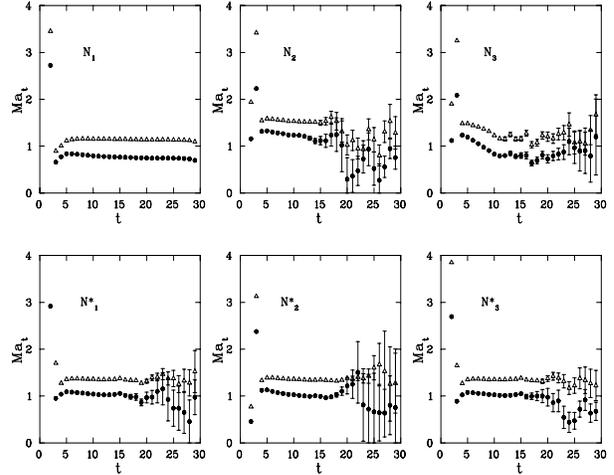}
\vspace{7cm}
\caption{Effective mass plots at two quark masses. }
\label{Emass_nstar_sl}
\end{figure}
\begin{figure}
\includegraphics{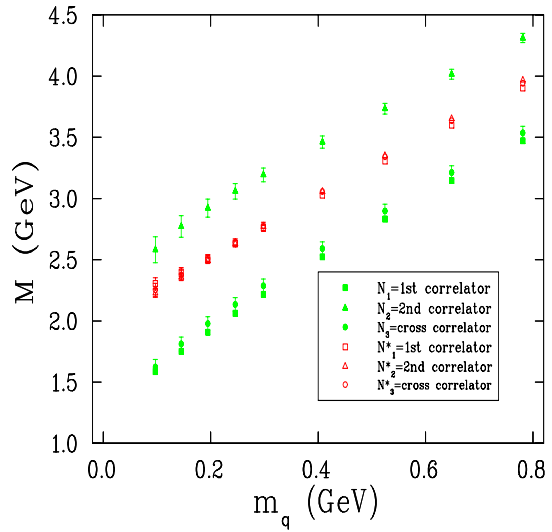}
\vspace{5.5cm}
\caption{Masses of $N_{1/2\pm}$ as a function of quark mass for all three types of
correlators.}
\label{QM_n6.ps}
\end{figure}
Figure~\ref{Emass_nstar_sl} shows some sample effective mass plots 
at quark mass no. 3 and 7 for all three correlators
in the source-sink combination of SL which has the best signal among LL,LS,SL,SS.
Valid plateaus exist in all cases.
Figure~\ref{QM_n6.ps} shows the extracted masses as a function of the quark mass.
The $N_1$ and $N_3$ give roughly the same mass, while $N_2$
yields a consistently higher mass.
It turns out that $\chi_2$ has very little overlap with the ground-state
nucleon and couples mostly to the first excited state, 
the Roper $N_{1/2+}(1440)$. 
This point will be clarified below.
All three negative-parity correlators give consistent masses.
The ordering of $N_{1/2+}(1440)$ and $N_{1/2-}(1535)$ at the quark masses
considered is inverted compared to that from experiment.
These results are consistent with those observed in~\cite{sas00}. 
It would be interesting to see if the ordering is reversed at smaller 
quark masses where meson cloud becomes more important. The crossing can be 
expected even in the quenched approximation where quarks wiggling backward in
time provide part of the meson cloud. Of course quenched QCD is sick at small
quark masses, so a correct description must include the sea quarks. 
There is already evidence of such crossing at very
small quark masses in the case of $a_0$ and $a_1$ mesons~\cite{hank00,dong00}.
The effects of the meson cloud show up as non-analytic behavior in the quark mass.
Some curvature is already apparent in the data so a linear extrapolation is 
undesirable.

\begin{figure}
\includegraphics{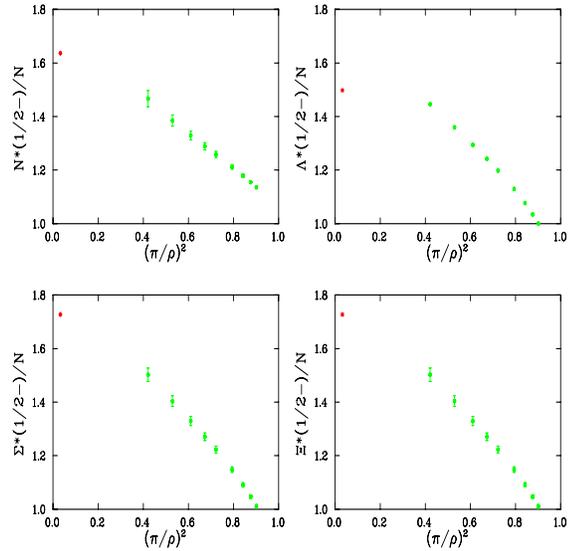}
\vspace{7cm}
\caption{Mass ratio N*/N as a function of $(\pi/\rho)^2$. The experiment
points are indicated on the left.}
\label{Ratio_nstar}
\end{figure}

In order to make contact with experiment, 
Figure~\ref{Ratio_nstar} presents mass ratios of the
$1/2-$ states to the nucleon ground state as a function of $(\pi/\rho)^2$.
This way of presenting data has minimal dependence on  the uncertainties 
in determining the scale and the quark masses.
These ratios appear headed in the right direction compared to experiment.
The $\Lambda_{1/2-}(1405)$ deserves special attention since it requires a
larger curvature.
At present it is almost degenerate 
with $N_{1/2-}(1535)$ while in the physical spectrum its mass is lower
despite having a heavier strange quark. This has been a long-standing puzzle.
The origin of the mass suppression of
the $\Lambda_{1/2-}(1405)$ lies in strong resonance channels such as the 
$KN$ and $\pi\Sigma$ which would present themselves as nonlinearities 
in the quark-mass extrapolation.
Whether it happens at smaller quark masses remains to be seen.
\begin{figure}
\includegraphics{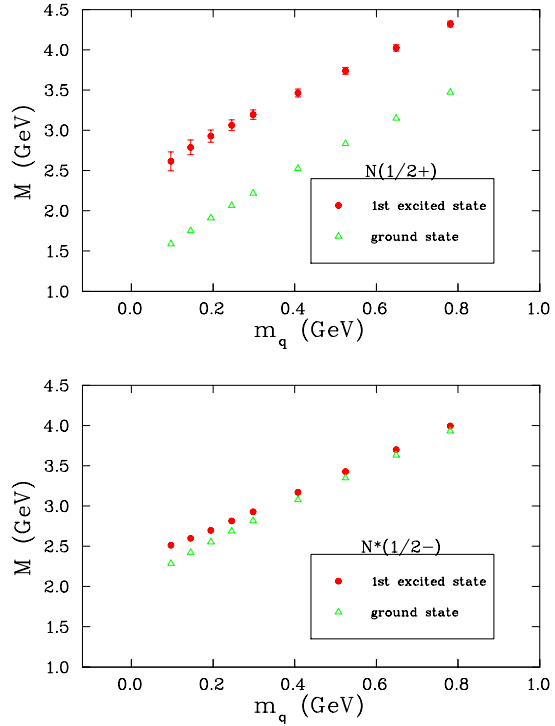}
\vspace{9.0cm}
\caption{Ground states and 1st excited states in the nucleon channel 
from the correlator matrix.}
\label{QM_one_n2}
\end{figure}

To extract information on the first excited states in each sector, 
the 2 by 2 
correlator matrix from the two interpolating fields is considered, using the
variational method originally proposed in~\cite{lus90}.
Figure~\ref{QM_one_n2} shows the results in the nucleon channel. 
It confirms that $\chi_2$ couples mostly to the the $N_{1/2+}(1440)$.
The splitting of the Roper $N_{1/2+}(1440)$ from the nucleon is 
roughly 5 times that of the $N_{1/2-}(1620)$ from the $N_{1/2-}(1535)$, 
consistent with experimental observations.
This may be traced back to the structure of the interpolating fields used to
excite the states. The $\chi_1$ and $\chi_2$ 
have upper spinor components of the order 1 and $(p/E)^2$ respectively,
whereas the interpolating fields for 1/2- states,
$\chi_1^-=\gamma_5\chi_1$ and $\chi_2^-=\gamma_5\chi_2$,
both have upper spinor components of the order $p/E$ which 
also introduce derivatives of the quark field operators.
In the quark model, the two nearby low-lying
$N_{1/2-}$ states of the physical spectrum are described by coupling
three quarks to a spin-1/2 or spin-3/2 spin-flavor wave function
coupled in turn to one unit of orbital angular momentum.  
In a relativistic field theory one does not expect $\chi_1^-$ and
$\chi_2^-$ to isolate individual states.  However, it is expected that
$\chi_1^-$ will predominantly excite the lower-lying $j=1/2$ state
associated with $\ell=1$ and $s=1/2$ while
$\chi_2^-$ will predominantly excite the higher-lying state associated
with $s=3/2$. This expectation is borne out in the variational analysis.
The results should improve with a larger basis of interpolating fields.
An independent method based on maximum entropy~\cite{mem} is 
being applied to verify the results.
\vspace*{-0.2cm}

\section{Conclusion} 
The potential of anisotropic lattices, smeared operators and correlator
matrix is demonstrated in probing the low-lying N*s in the $1/2+$ 
and $1/2-$ sectors.
The pattern of the splittings is mostly consistent with experiment.
More definite comparisons should address the systematics:
finite $a$, finite volume, statistics,
and most importantly, the chiral limit. 
There are still unresolved issues, such as the inverted ordering 
of the $N_{1/2+}(1440)$ and
the $N_{1/2-}(1535)$, and the degeneracy of the $\Lambda_{1/2-}(1405)$ and
the $N_{1/2-}(1535)$ at present quark masses,
that depend on this limit.
One way to proceed is to use ChPT to extrapolate the lattice data 
at quark masses where the quenched approximation is benign,
like that proposed in~\cite{lein99}. 
The more desirable way is to be able to simulate directly at or near
the physical quark masses. Toward this end, good chiral properties are
essential, and the domain-wall and the overlap quarks~\cite{neu00,dong99}
seem to hold the promise.

\section{Acknowledgment}
This work is supported in part by U.S. DOE under grant DE-FG02-95ER40907.
A combination of computing resources have been used,
including an allocation from NERSC, 
the Calico Alpha cluster at JLab, 
and the Origin 2000 at the Virginia Campus of GW.
I am indebted to D. Leinweber for initial collaborations on the
subject and continuing discussions. 
I also benefited from discussions with K.F. Liu, D. Richards, 
C. Morningstar, R. Fiebig, N. Isgur, R. Edwards.


\begin{thebibliography}{99}

\bibitem{cppacs00} 
S. Aoki, {\it et al.}, Phys. Rev. Lett. {\bf 84}, 238 (2000),
and these proceedings.

\bibitem{pdg00} 
Particle Data Group, Eur. Phys. J. C {\bf 15}, 1 (2000).

\bibitem{lee99} 
F.X. Lee, D.B. Leinweber, Nucl. Phys. B (Proc. Suppl.) {\bf 73}, 258 (1999).

\bibitem{sas00} 
S. Sasaki, Nucl. Phys. B (Proc. Suppl.) {\bf 83}, 206 (2000);
hep-ph/0004252;
 T. Blum, S. Sasaki, hep-lat/0002019;
 T. Blum, S. Ohta, S. Sasaki, these proceedings;

\bibitem{dgr00} 
D. Richards, these proceedings.

\bibitem{lein91} 
D.B. Leinweber, R.M. Woloshyn, T. Draper,
Phys. Rev. D {\bf 43},  1659 (1991).

\bibitem{mor97} 
C.J. Morningstar, M. Peardon, Phys. Rev. D {\bf 56}, 4043 (1997).

\bibitem{alf98} 
M. Alford, T.R. Klassen, G.P. Lepage, Phys. Rev. D {\bf 58}, 034503 (1998);
Nucl. Phys. {\bf B496}, 377 (1997).

\bibitem{lewis00} 
R. Lewis, these proceedings;

\bibitem{wolo00} 
R.M. Woloshyn, Phys. Lett. {\it B476}, 309 (2000).

\bibitem{rbc00} 
T. Manke, these proceedings.

\bibitem{Gus90} S. G\"{u}sken,
Nucl.Phys. (Proc. Suppl.) {\bf B17}, 361 (1990)

\bibitem{hank00} 
H. Thacker, private communication.

\bibitem{dong00} 
S.J. Dong, F.X. Lee, K.F. Liu, J.B. Zhang, these proceedings.

\bibitem{lus90} 
M. L\"{u}scher, U. Wolff,  Nucl. Phys. {\bf B339}, 222 (1990).

\bibitem{mem} 
M. Asakawa, T. Hatsuda, Y. Nakahara, hep-lat/9909137, hep-lat/0011040.

\bibitem{lein99} 
D.B. Leinweber, A.W. Thomas, K. Tsushima, S.V. Wright,
Phys. Rev. D {\bf 61},  074502 (2000).

\bibitem{neu00}
H. Neuberger, Nucl. Phys. B (Proc. Suppl.) {\bf 83-84}, 67 (2000); 
Phys. Lett. B {\bf 417}, 141 (1998).

\bibitem{dong99} 
S.J. Dong, F.X. Lee, K.F. Liu, J.B. Zhang,
Phys. Rev. Lett. (in press).

\end{thebibliography}
\end{document}